\documentclass[mote]{aa}
\usepackage{graphicx}
\usepackage{txfonts}
\begin{document}
\title{The detection of the ($J,K$)=(18,18) line of NH$_3$
        \thanks{Based on observations with the 100-m telescope 
        of the MPIfR (Max-Planck-Institut f{\"u}r
        Radioastronomie) at Effelsberg.}}

\subtitle{}

\author{T.~L.~ Wilson\inst{1,2}
        \and
        C.~Henkel\inst{1}
        \and
        S.~H{\"u}ttemeister\inst{3}}

\offprints{T.~L.~Wilson, e-mail: twilson@eso.org}

\institute{Max-Planck-Institut f{\"u}r Radioastronomie, Auf dem H{\"u}gel 69, 53121 Bonn, Germany
           \and
           European Southern Observatory, Karl-Schwarzschild-Stra{\ss}e 2, 85748 Garching, Germany
           \and
           Astronomisches Institut der Ruhr-Universit{\"a}t, D-44780 Bochum, Germany}

\date{Received May 12, 2006; accepted August 10, 2006}

% \abstract{}{}{}{}{} 
% 5 {} token are mandatory
 
\abstract
{}
{A study of the metastable ($J,K$)=(18,18) line of ammonia (NH$_3$) is presented aiming at better defining the physical properties of 
molecular gas in extreme environments.}
{The spectra were collected with the Effelsberg 100-m telescope and are analyzed in combination with other data using a rotation diagram 
(Boltzmann plot).}
{The first astronomical detection of the metastable ($J,K$) = (18,18) line of NH$_3$ is reported. This is the NH$_3$ line with by far the 
highest energy, 3130\,K above the ground state, detected in interstellar space. It is observed in absorption toward the galactic center star 
forming region Sgr~B2. There is a clear detection toward Sgr~B2~(M) and a likely one toward Sgr~B2~(N). An upper limit for emission is 
determined for Orion~KL. If we combine the (18,18) line results from Sgr~B2~(M) with the previously measured (12,12) absorption line, we 
find a rotation temperature, $T_{\rm rot}$, of $>$1300\,K for the absorbing cloud. This is at least a factor of two higher than previously 
derived values from less highly excited ammonia lines. $T_{\rm rot}$ gives a lower limit to the kinetic temperature. There is a hot low 
density gas component in the envelope of Sgr~B2. It is possible that the (18,18) line arises in this region. The radial velocity of the 
low density, hot envelope is the same as that of the dense hot cores, so the (18,18) line could also arise in the dense hot cores where 
non-metastable ($J$$>$$K$) absorption lines from energy levels of up to $\sim$1350\,K above the ground state have been observed. A 
discussion of scenarios is presented.}
{}


   \maketitle
%
%________________________________________________________________

\section{Introduction}

A number of vibrationally or torsionally excited molecules have been detected in the interstellar medium, among them H$_2$, CS, HCN, 
NH$_3$, SiO, H$_3C$N, CH$_3$CN, CH$_3$OH, C$_2$H$_3$CN and CH$_3$CH$_2$CN (e.g. Menten et al. 1986; Ziurys \& Turner 1986; Mauersberger 
et al. 1988; Hauschildt et al. 1993; Nummelin \& Bergman 1999; Mehringer et al. 2004). These lines often trace the hottest gas, where 
molecules can still survive, and thus serve as useful probes of the physical conditions of star forming regions. Particularly attractive 
are `hot core' environments, with temperatures and densities in excess of 100\,K and 10$^5$\,cm$^{-3}$, respectively. Here young massive 
stars heat the surrounding dust and gas and modify the chemical composition of the latter. 

There are, however, also lines in the vibrational ground state, matching or even surpassing the excitation requirements for some of the 
vibrationally excited transitions seen in hot cores. With the Kuiper Airborne Observatory, Watson et al. (1980) detected the CO $J$=21--20 
and 22--21 lines in Orion~KL, corresponding to an energy of $\sim$1200\,K. Observations of ammonia (NH$_3$) with the Effelsberg 100-m radio 
telescope yielded detections to the ($J,K$)=(14,14) inversion transition toward Orion~KL and Sgr~B2. These lines arise from energy levels 
$\sim$1950\,K above the ground state (H{\"u}ttemeister et al. 1995). 

Ammonia provides an excellent possibility to observe highly excited transitions since inversion transitions ($\Delta J$=0, $\Delta K$=0) 
across a ($J,K$) doublet are located in a band ranging from about 20 to 50\,GHz. Ammonia is one of the most interesting tracers of 
the physical properties of dense interstellar gas. It is an excellent diagnostic of kinetic temperature, $T_{\rm kin}$, over a wide range 
of densities. To a lesser extent, NH$_3$ is a probe of H$_2$ density, $n$(H$_2$). Inversion lines arise from either non-metastable 
($J$$>$$K$) or metastable ($J$=$K$) levels. The non-metastable levels can decay via allowed rotational transitions in the infrared (IR) 
with ($\Delta J$=$\pm$1, $\Delta K$=0) and have very short lifetimes. Thus, non-metastable lines require exceptional excitation 
conditions such as very large densities or intense IR radiation fields. In contrast, metastable ($J$=$K$) inversion lines arise from ($J,K$) 
doublet levels at the bottom of a $K$ ladder, so cannot rapidly decay to other levels. Radiative transitions between metastable levels are 
first order forbidden, so the populations of these levels are exchanged by collisions. By combining column densities of different metastable 
levels, one can obtain an estimate of $T_{\rm rot}$, which is a lower limit to $T_{\rm kin}$. The value of the excitation temperature, 
$T_{\rm ex}$, across an inversion doublet will be one half of $T_{\rm kin}$ when $n$(H$_2$) $\sim$ 10$^4$\,cm$^{-3}$ (see e.g. Rohlfs \& 
Wilson 2003 for details). 

In this paper, we report the first detection of the (18,18) inversion line, toward Sgr~B2~(M). We also present a likely detection 
toward Sgr~B2~(N) and an upper limit for Orion~KL. We interpret the data using comparisons with previous results involving lines of lower 
excitation.

\section{Observations}

All data were taken with the 100-m radio telescope of the MPIfR at Effelsberg on July 24 and 26, 1998. At the line frequency, $\nu_{18,18}$ 
= 46.123297\,GHz, the Full Width to Half Power (FWHP) beamsize is (21$\pm$2)\arcsec, determined from observations toward PKS\,1830--211
(typical elevations 15$^{\circ}$--18$^{\circ}$) and 3C\,273 (30$^{\circ}$--35$^{\circ}$). A single polarisation HEMT receiver ($T_{\rm rec}$
$\sim$ 70\,K) at the primary focus provided system temperatures of 250--500\,K on a $T_{\rm A}^*$ scale with a characteristic sensitivity 
of $\sim$0.5\,K/Jy and beam and aperture efficiencies of 33\% and 16\%, respectively. The backend was an autocorrelator ("AK90") with 
eight spectrometers, each containing 512 channels and covering 40\,MHz. The channel separation was 0.5\,km\,s$^{-1}$.
 
Spectral line measurements were interspersed with continuum measurements. The procedure was first to carry out a pointing cross
scan on Sgr~B2~(M), correct the telescope pointing, and then measure a number of spectra. Spectra were measured by first taking data 
at a position at an offset of --3$^{\rm m}$ in Right Ascension for 3 minutes, then repeating this procedure toward the source. For 
each scan, the final spectrum is the difference between on- and off-source measurement divided by the off-source measurement. For 
a number of scans, we shifted the radial velocities to make certain that the lines found are not caused by instrumental effects. 

Sgr~B2 rises to a maximum elevation of 11$^{\circ}$, so for this source calibration of spectral line data was accomplished by measurements 
of its strongest continuum source, Sgr~B2~(M). With an observed FWHP of (23$\pm$2)\arcsec\, the source appears to be slightly resolved. 
Adopting a 46.12\,GHz continuum flux density of $\sim$20\,Jy, measured at 43\,GHz with a FWHP of 39\arcsec\ by Akabane et al. (1988), 
assuming that the entire flux density is within our beam, and accounting for the difference between beam size and FWHP of the source,
the resulting peak continuum temperature becomes $T_{\rm C}$ $\sim$ 25\,K in units of main beam brightness temperature. This peak continuum
temperature was used to determine the line brightness temperature for Sgr~B2~(M) from our measured line-to-continuum ratios. For Sgr~B2~(N), 
we have no reliable continuum measurement, so a continuum flux density ratio of Sgr~B2~(M) to (N), $\sim$1.8 (e.g. Akabane et al. 1988; 
H{\"u}ttemeister et al. 1993), was used to determine its peak main beam brightness temperature. The absolute calibration uncertainty of 
the continuum data is about $\pm$20\% for Sgr~B2~(M) and $\pm$30\% for Sgr~B2~(N). For the line data, see Sect.\,3.

Orion~KL rises to a maximum elevation of 34$^{\circ}$ at Effelsberg. Pointing scans were made toward PKS\,0420-01 and 3C\,161, but there 
is no nearby calibrator. For calibration we have therefore used NGC\,7027, assuming a flux density of 5\,Jy (Ott et al. 1994), correcting 
for the source size of 7\arcsec$\times$10\arcsec (also Ott et al. 1994) and accounting for differences in elevation applying the 
formula\footnote{see {\it www.mpifr-bonn.mpg.de/div/effelsberg/calibration/calib.html}} given by A. Kraus. For Orion~KL, the calibration 
uncertainty is of order $\pm$15\%.

A search of the JPL molecular line catalog showed that there were only very rare species within 10 MHz (=65 km s$^{-1}$) of the (18,18) 
transition. Given the intensity of the lines (see Table 1), we conclude that there is no accidental line overlap.

\begin{figure}[ht]
\label{data}
{\includegraphics[scale=0.70,angle=-90]{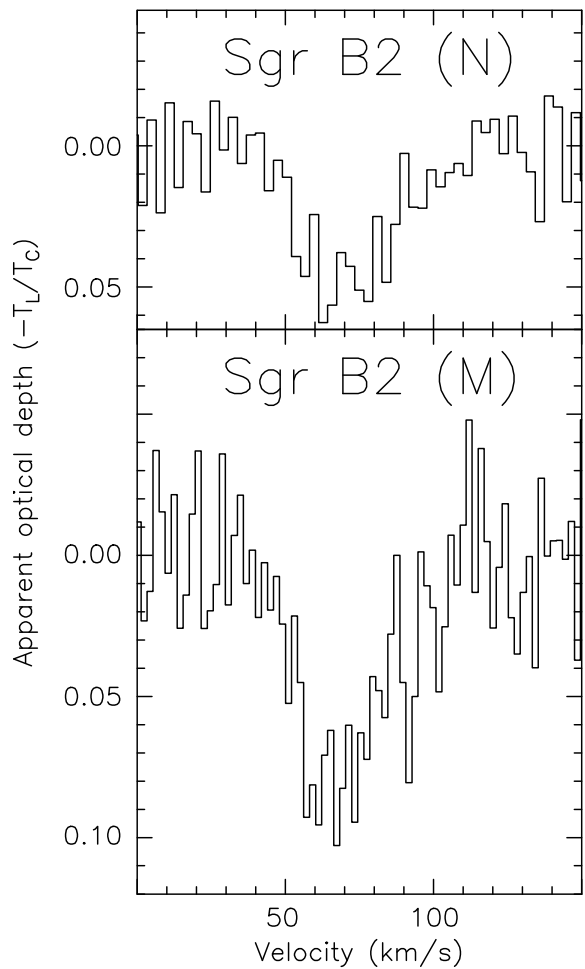}} 
\caption{Spectra from Sgr~B2~(N) (upper panel) and Sgr~B2~(M) (lower panel), taken on July 24, 1998, in terms of apparent optical depth, 
$\tau_{\rm app}$. The (absolute) line intensity is small compared to the continuum intensity, so the apparent optical depth is approximately 
the line to continuum ratio, $|T_{\rm L}|$/T$_{\rm C}$. Channel widths are 3.05\,km\,s$^{-1}$ and 2.03\,km\,s$^{-1}$ for the upper and 
lower spectrum, respectively. Baselines of order one and two were subtracted. For gaussian fit parameters, see Table 1.}
\end{figure}

\section{Results}

As already indicated (Sect.\,1), the extensive NH$_3$ data sets of H{\"u}ttemeister et al. (1993, 1995) show absorption in the metastable 
inversion lines of Sgr~B2 up to the (14,14) line. Thus we would also expect absorption in the (18,18) metastable line. The profiles in 
Fig.\,1 show good agreement with the line parameters $V_{\rm LSR}$ and $\Delta V_{1/2}$ of previous data. We claim a definite detection 
toward Sgr~B2~(M), where there is a good signal-to-noise ratio, and a likely detection toward Sgr~B2~(N), while Orion~KL remains undetected. 
At $\sim$3130\,K above the ground state (the energy difference between the two states of the inversion doublet is only 2.2\,K), the (18,18)
transition is by far the highest excited ammonia line ever detected in the interstellar medium. The previously measured (14,14) lines in
the vibrational ground state and the (2,1)--(1,1) transition in the vibrationally excited $\nu_2$ state (Wilson et al. 1993; H{\"u}ttemeister 
et al. 1995; Mauersberger et al. 1988) are, with $\sim$1945 and 1400\,K, located far below the (18,18) inversion doublet reported here. 

Only a small fraction of the continuum background is absorbed by the ($J,K$) = (18,18) lines so that, assuming a source covering factor 
of unity, the lines are optically thin and optical depths can be directly obtained from the line-to-continuum ratio, $|T_{\rm L}|$/$T_{\rm C}$. 
Sgr~B2~(M) and (N) were observed twice (Sect.\,2). Line intensity ratios between the two sources are 2.9$\pm$0.7 and 2.7$\pm$1.1 for the first 
and second measurement, respectively. The good agreement between these two ratios as well as a continuum flux density ratio of $\sim$1.8 
(Sect.\,2) then suggest that the optical depth of the (18,18) line is higher toward Sgr~B2~(M). There is not only excellent agreement between 
Sgr~B2~(M) to (N) line intensity ratios. Radial velocities and linewidths of Sgr~B2~(M) and (N) are also almost identical for each observing 
epoch. Comparing, however, the spectra from the first with those of the second observing period, there are striking differences. The data 
taken on July 26, 1998, show amplitudes that are lower by factors of 2--3, velocities that are lower by 5\,km\,s$^{-1}$ and linewidths that 
are reduced by 40\% (Table~1). While this is readily explained by pointing problems affecting the July 26 data (for a point source, this would 
correspond to an offset of 12\arcsec), we will nevertheless use both results, (1) to derive a definite lower limit of $T_{\rm rot}$ in Sect.\,4 
and (2) to more thoroughly discuss the nature of the source in Sect.\,5. Fig.\,1 shows the spectra obtained on July 24, 1998.

\begin{table}
\caption{NH$_3$ ($J,K$)=(18,18) line parameters$^{\rm a)}$}
\begin{flushleft}
\label{lineparameters}
\begin{tabular}{l c l c c c}
\hline
\multicolumn{1}{c}{(1)} & 
\multicolumn{1}{c}{(2)} & 
\multicolumn{1}{c}{(3)} & 
\multicolumn{1}{c}{(4)} & 
\multicolumn{1}{c}{(5)} & 
\multicolumn{1}{c}{(6)} \\ 
\multicolumn{1}{c}{Source} & 
\multicolumn{1}{c}{$T_{\rm C}$} & 
\multicolumn{1}{c}{$|$$T_{\rm L}$$|$} & 
\multicolumn{1}{c}{$\tau_{\rm app}$} & 
\multicolumn{1}{c}{$V_{\rm LSR}$} & 
\multicolumn{1}{c}{$\Delta V_{1/2}$} \\
\multicolumn{1}{c}{} & 
\multicolumn{2}{c}{(K)} & 
\multicolumn{1}{c}{} & 
\multicolumn{2}{c}{(km\,s$^{-1}$)} \\ 
\hline
                         &                &               &                 &            &               \\
Sgr~B2~(M)$^{\rm b)}$    &     25         & 2.0$\pm$0.4   & 0.080$\pm$0.015 & 69$\pm$2   & 33$\pm$4      \\
                         &                & 0.8$\pm$0.2   & 0.031$\pm$0.007 & 64$\pm$2   & 19$\pm$3      \\
Sgr~B2~(N)$^{\rm b)}$    &     14         & 0.7$\pm$0.1   & 0.050$\pm$0.009 & 71$\pm$2   & 33$\pm$4      \\
                         &                & 0.3$\pm$0.1   & 0.020$\pm$0.007 & 66$\pm$2   & 20$\pm$5      \\
Orion~KL$^{\rm c)}$      &                & $\le$0.1      &       ---       &    ---     &    ---        \\
                         &                &               &                 &            &               \\
\hline
\end{tabular}
\ \ \ \ \ \ \ \\
\item[a)] Continuum ($T_{\rm C}$) and line temperatures ($T_{\rm L}$) are given on a main beam brightness temperature scale ($T_{\rm MB}$;
for details, see Sect.\,2). $\tau_{\rm app}$ denotes the apparent optical depth (for a definition, see the caption to Fig.\,1), $V_{\rm LSR}$
marks the Local Standard of Rest velocity and $\Delta V_{1/2}$ gives the Full Width to Half Power (FWHP) linewidth.  
\item[b)] Data taken on July 24, 1998 (upper line), and July 26, 1998 (lower line). Among the two sets of gaussian parameters, we consider 
the second  to be affected by pointing problems.  
\item[c)] The 3$\sigma$ limit for a single 0.5\,km\,s$^{-1}$ wide channel is 0.2\,K on a $T_{\rm MB}$ scale. The expected emission line 
at $\sim$5.5\,km\,s$^{-1}$ should have a width of $\sim$8\,km\,s$^{-1}$ (Wilson et al. 1993). Thus, after averaging several channels, our 
3$\sigma$ sensitivity becomes 50--100\,mK.
\end{flushleft}
\end{table}

\section{Determination of rotation temperatures}

The NH$_3$ populations are divided into ortho and para species. Only the levels with $K$=0, 3, 6, 9, 12, 15, 18, etc., belong to ortho-NH$_3$. 
While the total number of para levels is twice that of ortho levels, the total abundance of the ortho and para species is equal in Local
Thermodynamical Equilibrium (LTE). A statistical weight of $g_{\rm o,p}$ = 2 for the ortho- and 1 for the para-levels accounts for this.
A special case is the (3,3) doublet inversion line. One half of the levels in the $K$=$0$ ladder are missing. As a consequence, in some cases, 
the (3,3) doublet populations are inverted. In a few sources, masers have been found in the transition connecting the two states of the (3,3) 
inversion doublet (for Sgr~B2, see Martin-Pintado et al. 1999). In the following we assume that the high lying ortho-NH$_3$ doublet populations 
($J$,$K$$\geq$12) do not deviate from LTE.

Since ortho and para populations cannot be interchanged by collisions with H$_2$, there can be deviations from their LTE abundance ratio of unity.
Thus we compare the (18,18) column densities with those of the (12,12) line, the previously observed metastable ortho-NH$_3$ transition nearest 
in energy, 1450\,K above the ground state. For our measurements, the relevant  equation relating apparent optical depth to column density of the 
doublet is  
$$
  \frac{N(J,K)}{T_{\rm ex}}  = 1.61\times 10^{14}\,\,\,\frac{J(J+1)}{K^2 \nu}\,\,\,\,\,\tau \,\,\,\Delta V_{1/2}.
$$
Here $N$ is in cm$^{-2}$, $T_{\rm ex}$ and $T_{\rm mb}$ in K, $\nu$ in GHz and $\Delta V_{1/2}$ in km\,s$^{-1}$. Using the (12,12) and 
(18,18) lines to estimate $T_{\rm rot}$ in a Boltzmann relation 
$$
 \frac{N(J_1,K_1)}{N(J_2,K_2)} = \frac{(2J_1 + 1)}{(2J_2 + 1)} \frac{g_{\rm o,p,1}}{g_{\rm o,p,2}}\,\,\,\,\,{\rm exp}(-{\rm (E_1 - E_2)}/T_{\rm rot})
$$
does not require any assumptions about ortho/para ratios. In Table~2 we list the column densities, the normalized column densities (i.e. column 
densities divided by the statistical weights, (2$J$+1)$\,g_{\rm o,p}$), and the resulting rotation temperatures. In Fig.\,2 we show for Sgr~B2~(M) 
a plot of normalized column densities per $K$ excitation temperature versus energy of the level above the ground state. Implicit in this fit is 
the assumption that $T_{\rm ex}$ is the same for all metastable transitions. This will be discussed in Sect.\,5.2.

\section{Discussion}

\subsection{The Sgr~B2 cloud}

The Sagittarius B2 molecular cloud is one of the most active regions of star formation in our Galaxy. It is located about 40\arcmin\ from 
the galactic center, at a projected distance of $\sim$100 pc for a Sun-galactic center distance of 8.5~kpc. Sgr~B2 hosts three prominent 
cores of ionized gas, Sgr~B2~(M), (N) and (S) (see e.g. Martin-Pintado et al. 1999). These are associated with molecular clouds which are in 
turn embedded in a much more extended, lower density envelope.

\begin{table}
\caption{Rotational temperatures$^{\rm a)}$}
\label{masses1}
\begin{flushleft}
\begin{tabular}{ l r r r r r }
\hline
\multicolumn{1}{c}{(1)}             &
\multicolumn{1}{c}{(2)}             &
\multicolumn{1}{c}{(3)}             &
\multicolumn{1}{c}{(4)}             &
\multicolumn{1}{c}{(5)}             &
\multicolumn{1}{c}{(6)}             \\
\multicolumn{1}{c}{Source}          &
\multicolumn{2}{c}{Column}          &
\multicolumn{2}{c}{Normalized}      &
\multicolumn{1}{c}{$T_{\rm rot}$}   \\
\multicolumn{1}{c}{}                &
\multicolumn{2}{c}{density}         &
\multicolumn{2}{c}{column}          &
\multicolumn{1}{c}{}                \\
\multicolumn{1}{c}{}                &
\multicolumn{2}{c}{}                &
\multicolumn{2}{c}{density}         &
\multicolumn{1}{c}{}                \\
\multicolumn{1}{c}{}                &
\multicolumn{1}{c}{(12,12)}         &
\multicolumn{1}{c}{(18,18)}         &
\multicolumn{1}{c}{(12,12)}         &
\multicolumn{1}{c}{(18,18)}         &
\multicolumn{1}{c}{}                \\
\multicolumn{1}{c}{}                &
\multicolumn{4}{c}{(cm$^{-2}$/K)}   &
\multicolumn{1}{c}{(K)}             \\
\hline
Sgr~B2~(M)  & 5.6(12) &    1.0(13) & 1.1(11) &    1.3(11) &$>$1300 \\
            &         &    2.2(12) &         &    3.0(10) &   1300 \\
Sgr~B2~(N)  & 1.5(13) &    6.2(12) & 2.9(11) &    8.4(10) &   1350 \\
            &         &    1.5(12) &         &    2.0(10) &    600 \\
Orion KL    & 2.6(13) & $<$3.0(12) & 5.2(11) & $<$4.1(10) & $<$500 \\
\hline
\end{tabular}
\ \ \ \ \ \ \ \\
\item[a)] The (12,12) data for Sgr~B2 were taken from Table~2 of H{\"u}ttemeister et al. (1995). As in Table~1 of this paper, the higher and 
lower (18,18) column densities refer to the data taken on July 24 and 26, 1998, respectively (see also Sect.\,3). Errors are not given, because 
a comparison between the two epochs provides a good measure of the extremes involved. For Orion~KL, the (12,12) column density was taken from 
Wilson et al. (1993). For the (18,18) column density optically thin emission, $T_{\rm mb}$=0.1\,K and a FWHP linewidth of $\Delta V_{\rm 1/2}$ = 
8\,km\,s$^{-1}$ was assumed. The given rotation temperature also accounts for the fact, that Wilson at al. calculated column densities averaged 
over a 29\arcsec\ beam, while the (18,18) upper limit is an average over a 21\arcsec\ beam.
\end{flushleft}
\end{table}

For Sgr~B2, the geometry of the molecular clouds and continuum sources is important. There are compact radio continuum sources absorbed by NH$_3$ 
inversion lines. The telescope beam of the Effelsberg 100-m dish is larger than these continuum sources and the associated hot cores, but smaller 
than the NH$_3$ containing cloud in the molecular envelope of Sgr~B2 (see e.g. H93 for details of this geometry and 40\arcsec\ images of NH$_3$ lines). 

For the NH$_3$ ($J,K$) = (1,1), (2,2), (3,3), (4,4), (5,5), (6,6) and (7,7) metastable inversion lines, there is absorption toward Sgr~B2~(M) 
and (N), and emission one full beamwidth, 40\arcsec, off the continuum sources. H93 assumed that this is due to a uniform extended layer of 
NH$_3$. In this case, the main beam brightness temperature of the continuum source, $T_{\rm C}$, is larger than the excitation temperature, 
$T_{\rm ex}$, of the line. Given a uniform NH$_3$ cloud, more extended than the radio telescope beam and calculated collision rates (Danby 
et al. 1988), one can determine values for $n$(H$_2$). From such models, H93 had estimated $n$(H$_2$)$\sim$3\,10$^3$\,cm$^{-3}$. Since radiative 
transitions across $K$ ladders are forbidden, the relation of $T_{\rm rot}$ is close to but less than $T_{\rm kin}$; this is 
not strongly dependent on H$_2$ densities. Thus, from measurements of a number of metastable NH$_3$ absorption lines, one estimates that 
$T_{\rm rot}$ = 160\,K in the envelope of Sgr~B2 (Fig.\,2). H{\"u}ttemeister et al. (1995) carried out measurements of the (8,8), (9,9), 
(10,10), (11,11), (12,12), (13,13), and (14,14) metastable inversion lines toward Sgr~B2~(M) and (N). For SgrB2(M), $T_{\rm rot}$$\sim$600\,K
(Fig.\,2). Toward the continuum sources these lines were found in absorption. Because of the weakness of the lines, there were no measurements 
of off-source positions, where emission corresponding to the absorption would be expected as with the (1,1) to (7,7) lines. 

The rotational transitions of NH$_3$ produce lines in the far infrared (FIR). An extensive study of these FIR lines has been carried out using 
the ISO (Infrared Space Observatory) LWS (Long Wavelength Spectrometer) with $\Delta \lambda$/$\lambda$ $\sim$ 8000 and a beam size of 
$\sim$80\arcsec\ (Ceccarelli et al. 2002). Twenty-one NH$_3$ lines in the wavelength range between 47 and 196$\mu$m were detected, all of them 
in absorption. Of these, 13 lines involve metastable levels, while 8 connect non-metastable levels. As Ceccarelli et al. (2002) point out, their 
lines absorb the spatially extended FIR dust continuum rather than the more compact free-free continuum absorbed by the centimeter wavelength 
inversion lines. This may account for some of the differences between inversion line results and the FIR line data. Ceccarelli et al. (2002) 
have presented a model for their data. This involves an absorbing layer 1.15$\pm$0.15\,pc in front of the FIR continuum source. This layer has 
a kinetic temperature of 700$\pm$100\,K and an H$_2$ density of $\le$10$^5$\,cm$^{-3}$. From the non-detection of FIR CO lines, Ceccarelli et 
al. (2002) surmise that $n$(H$_2$) $\le$ 10$^4$\,cm$^{-3}$. Oka et al.~(2005) deduce H$_2$ densities of order 10$^2$\,cm$^{-3}$ in the galactic 
center region, so the envelope of SgrB2 may have even lower densities. 

As with the other extended molecular clouds in the galactic center region, dust in the envelope of Sgr~B2 is comparatively cold, with dust 
temperatures, $T_{\rm dust}$, of order $\sim$70\,K. Thus in the envelope of Sgr~B2, non-metastable levels of NH$_3$ cannot be populated by 
collisions or IR fields. In addition to the envelope, however, there are a number of hot, dense, compact cores. For SgrB2(M), these have the 
same radial velocities as the envelope. In the cores, the populations of non-metastable levels can be significant. From maps of the (2,1), 
(3,2), (4,3), (5,4) and (4,2) non-metastable inversion emission lines, H93 found that the {\it warm envelope} has a FWHP size of $\sim$1\arcmin. 
These lines arise from energy levels up to 300\,K above the ground state. Measurements of non-metastable lines from energy levels between 400 and 
700\,K above the ground state show a mixture of absorption and emission. Above 700\,K, there is only absorption. 

H93, de Vicente et al. (1997), Ceccarelli et al. (2002) and Comito et al. (2003) proposed scenarios for Sgr~B2 including (1) compact {\it 
hot cores} with $T_{\rm kin}$$\sim$200\,K and $n$(H$_2$)$\sim$3\,10$^7$\,cm$^{-3}$, (2) a more extended {\it warm envelope} with 
$T_{\rm kin}$$\sim$40\,K and $n$(H$_2$)$\sim$3\,10$^5$\,cm$^{-3}$, and (3) a very extended {\it hot envelope} with $T_{\rm kin}$$\sim$200\,K 
and $n$(H$_2$)$<$10$^4$\,cm$^{-3}$. In view of the metastable NH$_3$ inversion line data of H{\"u}ttemeister et al. (1995) and the FIR line 
data of Ceccarelli et al. (2002), the $T_{\rm kin}$ of at least a part of the hot envelope should be raised to 600--700\,K.

\begin{figure*}[ht]
\label{boltzmann}
%\resizebox{\hsize}{!}
{\includegraphics[scale=0.70]{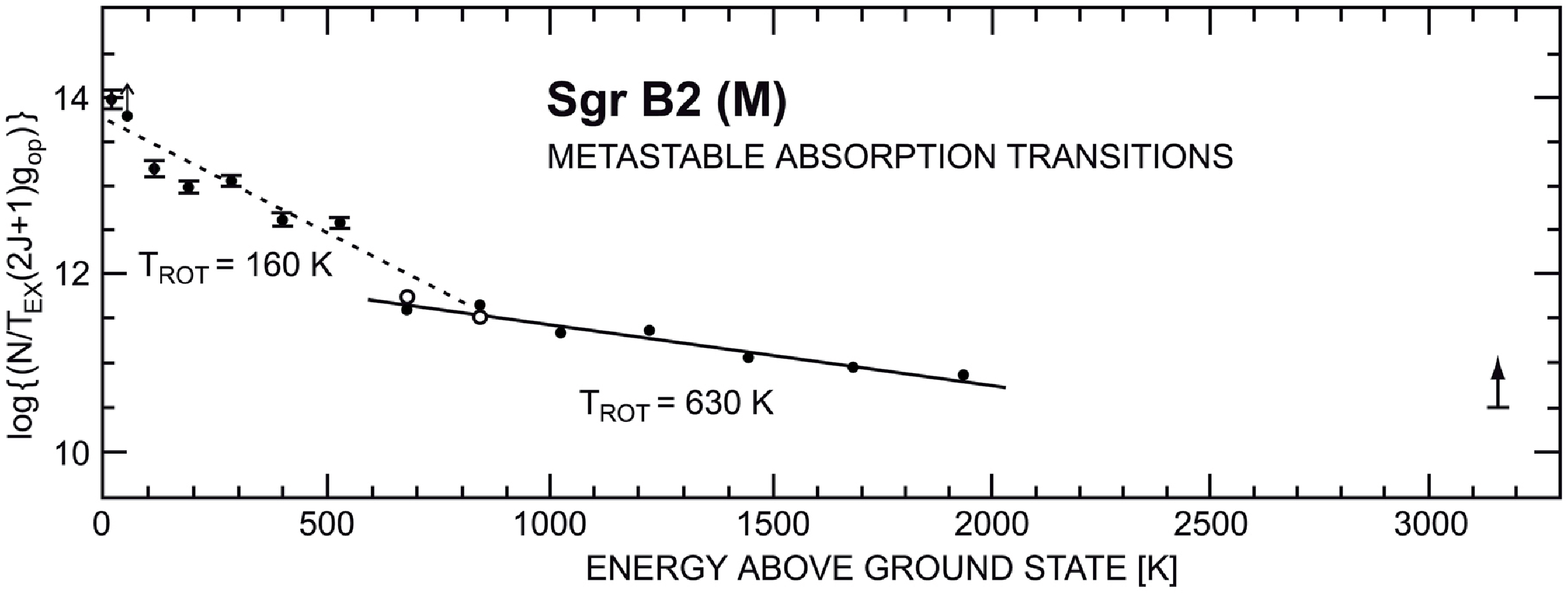}} 
\caption{Normalized column density per K excitation temperature ($T_{\rm ex}$) versus energy above the ground state. Filled circles with
error bars mark the ($J,K$)=(1,1)--(7,7) transitions measured by H{\"u}ttemeister et al. (1993), empty circles refer to the (8,8) and (9,9) 
lines observed by Wilson et al. (1982) and filled circles show the normalized column densities of the (8,8) to (14,14) transitions obtained 
by H{\"u}ttemeister et al. (1995). The inverse slope of a line connecting data points is a direct measure of the rotation temperature, 
$T_{\rm rot}$, as long as Local Thermodynamical Equilibrium can be assumed. While the concave form of the diagram suggests the presence of 
three gas layers with different temperature ($T_{\rm rot}$ = 160\,K, 630\,K and $>$1300\,K), the gas {\it might} actually be isothermal with 
$T_{\rm rot}$ approaching $T_{\rm kin}$ with increasing energy above the ground state (see the radiative transfer calculations of Walmsley 
\& Ungerechts (1983), Danby et al. (1988), and Flower et al. (1995)). Collision rate coefficients (Danby et al. 1988) were computed
for levels as high as ($J,K$) = (6,6) and temperatures as large as 300\,K. For higher transitions and temperatures, extrapolated rate
coefficients have to be used.}
\end{figure*}

\subsection{Relevance of the ($J,K$)=(18,18) line}

The most definite new result is the detection of the ($J,K$)=(18,18) line in absorption toward Sgr~B2~(M). There are two possible 
interpretations: 

\noindent
(1) As with the (1,1) to (7,7) lines, the (18,18) line arises in the extended envelope and is sub-thermally excited, 
or 

\noindent
(2) the (18, 18) line is thermalized and arises in a region close to the H{\sc ii} regions in Sgr~B2. After discussing the assumption 
of LTE used in Table~2 and Fig.\,3, these scenarios will be analyzed in more detail. 

With $T_{\rm kin}$ $>$ $T_{\rm rot}$ $>$ 1300\,K (Table~2), the ($J,K$) = (18,18) line reveals a molecular gas component with a 
kinetic temperature clearly exceeding the 600--700\,K estimated by H{\"u}ttemeister et al. (1995) and Ceccarelli et al. (2002). 
Before reaching a conclusion, however, we should ask whether the $T_{\rm ex}$ values of the various metastable inversion lines are really 
the same. Does the assumption of LTE hold? One can use the off-peak line and continuum results to show that the value of $T_{\rm ex}$ 
for the (1,1) line is $\sim$6\,K. A calculation using a two level model for metastable levels with a constant collision rate and the Einstein 
A coefficients shows that the excitation temperatures across metastable levels decrease from 6\,K for the (1,1) levels to slightly more 
than 3\,K for the (18,18) levels. Our fit to the normalized column densities in Fig.\,2 is based on the assumption that the value of 
$T_{\rm ex}$ is the same for each line. If the value of $T_{\rm ex}$ does follow the simple relation discussed previously, the 
normalized column densities will decrease more steeply than shown in Fig.\,3 with increasing $J$ value. This will lead to a lowering 
of the $T_{\rm rot}$ values obtained (Table~2; Fig.\,3) by about 20\%; however $T_{\rm rot}$ will still be $\ga$1000\,K. 

In Fig.\,2, there is a clear gradient in the values of $T_{\rm rot}$. Wilson et al.~(1993) found a similar effect for the high density 
region Orion~KL. This was interpreted as a gradient in $T_{\rm kin}$. We note, however, that in a warm enviroment radiative transfer 
simulations predict higher $T_{\rm rot}$ values, gradually approaching $T_{\rm kin}$, for higher metastable levels even if the gas is 
characterized by a single kinetic temperature (see Walmsley \& Ungerechts 1983; Danby et al.~1988; Flower et al.~1995). The situation 
for SgrB2~(M) may be more complex than that for Orion~KL. The lower energy ($J$$\le$7) metastable absorption lines shown in Fig.\,2 
definitely arise from the envelope, but the higher lying lines {\it might} arise in hot cores.

There are several arguments suggesting that the ($J,K$)=(18,18) emission arises in a warm, extended low density envelope:

\noindent
(1) Absorption
with similar apparent optical depths is seen from the lowest to the highest metastable inversion lines toward both Sgr~B2~(M) and (N). 
This includes the (18,18) line. Not a single other galactic star formation region shows a similarly excited NH$_3$ absorption component
so that statistically it appears unlikely that Sgr~B2 hosts two such clouds. With Sgr~B2~(M) and (N) being separated by only 
$\sim$47\arcsec\ ($\sim$2\,pc), it is not farfetched to assume that the absorption is caused by a single cloud with an extent of at 
least an arcminute. 

\noindent
(2) Maps made in the lower ($J,K$) transitions up to the (7,7) line (see Sect.\,5.1) reveal emission surrounding 
the radio continuum sources that appears to arise from the same low density gas component as the gas measured in absorption. Lack of evidence 
for such emission in $J$$>$7 metastable lines does not argue against this scenario, since such emission is weak and would be difficult to 
detect. 

\noindent
(3) Galactic disk molecular clouds extending over $\sim$2\,pc are commonly characterized by a low density. The cloud size-density 
relation of Larson (1981; his eq.\,5) suggests $n$(H$_2$)$\sim$10$^3$\,cm$^{-3}$. Although an extension of this relation for the galactic 
disk to galactic center clouds with their larger linewidths may not be justified, this is exactly the density that is required for 
absorption to dominate in a 30\arcsec\ -- 40\arcsec\ beam (see Sect.\,5.1) that characterizes most ammonia measurements obtained at Effelsberg. 

These arguments lead one to associate the (18,18) line with warm, extended, low density gas. However a scenario favoring a hot core origin 
cannot be excluded:

\noindent
(1) $T_{\rm C}$, averaged over a few arcseconds only, is of order $\sim$1000\,K (for detailed continuum maps, 
see e.g. Gaume \& Claussen 1990; De Pree et al. 1998). Thus the absorption may arise from compact hot regions of high density 
($T_{\rm ex}$$>$100\,K) directly associated with the hot cores.

\noindent
(2) Radial velocities of the dense, compact hot cores and less compact 
lower density envelope are nearly the same and it is not possible to separate their contributions kinematically. 

\noindent
(3) While it is sometimes argued that the lack of absorption in the ($J,K$) = (3,2), (4,3) and (7,6) VLA data of Vogel et al. (1987) excludes 
any compact high density absorbing cloud, the single-dish data of H93 show that absorption dominates the highly excited non-metastable transitions 
including the (7,6) line toward Sgr~B2~(M). These lines must arise (see Sect.\,5.1) from small regions that can only be slightly more extended than 
the radio continuum emission. 

\noindent
(4) A comparison of our data from July 24 and 26 shows systematic differences in the Sgr~B2~(M) and (N) absorption profiles (Sect.\,3), affecting 
more the high than the low velocity part of the spectra. This suggests that lower ($\sim$65\,km\,s$^{-1}$) and higher ($\sim$80\,km\,s$^{-1}$) 
velocity components are not smoothly distributed. 

A number of heating sources have been proposed for the galactic center region. These include PDRs (Photon Dominated or Photo Dissociation 
Regions; e.g. Goicochea et al.  2004), shocks (e.g. Flower et al. 1995; Ceccarelli et al. 2002), XDRs (X-ray Dominated Regions; e.g. 
Mart\'in-Pintado et al. 2000), cosmic rays (e.g. G{\"u}sten at al. 1981) and the dissipation of supersonic cloud motions (e.g. Wilson 
et al. 1982; Hasegawa et al. 1994). In Sgr~B2, several of these mechanisms are at work. Since kinetic temperatures in excess of 1000\,K 
are not widespread ($\gg$2\,pc), the latter two mechanisms will not be dominant for the gas traced by the NH$_3$ ($J,K$) = (18,18) line. 
Due to its low threshhold for photoionization ($\sim$4.1\,eV; Suto \& Lee 1983) NH$_3$ is rapidly destroyed by UV radiation and PDRs 
can also be excluded (for NH$_3$ in the `PDR galaxy' M\,82, see Wei{\ss} et al. 2001; Mauersberger et al. 2003). While detailed models for 
NH$_3$ being exposed to a strong X-ray radiation field have yet to be reported, shocks are widespread in the galactic center region
(e.g. H{\"u}ttemeister et al. 1998), can release NH$_3$ from grain mantles and can heat the gas to $T_{\rm kin}$$\sim$1000\,K and more (see 
Rizzo et al. 2001 for NH$_3$ in a cloud apparently being shocked by the wind of a Wolf-Rayet star). Mart\'in-Pintado et al. (1999) have 
found evidence for ring like structures in NH$_3$ in the envelope of Sgr~B2. It may be that the (18,18) emission could arise from such 
structures, although the $T_{\rm rot}$ values obtained from lines with $J$$\leq$4 are less than 200\,K. 

Shocks imply that large amounts of water vapor should also be produced. Simulations using shock models can reproduce the observed NH$_3$ column 
densities up to the (14,14) line (Flower et al.~1995; their Table~4). These models also predict water vapor column densities of $N$(H$_2$O) $\sim$ 
10$^{18}$\,cm$^{-2}$ for the envelope. From H$_2^{18}$O absorption line measurements and radiative transfer calculations, Comito et al. (2003) 
derive $N$(H$_2$O) = 3.5\,10$^{16}$\,cm$^{-2}$ in their Hot Layer, which they identify with a part of the Sgr~B2 envelope (for earlier measurements, 
see Neufeld et al. 2000; Ceccarelli et a. 2002). This is $\sim$30 times less than the theoretical prediction. Removing the gas seen in the (18,18) 
absorption line and a part of the gas giving rise to the lower metastable inversion lines from the envelope and associating it instead with 
the hot cores would reduce the required H$_2$O column density in the envelope and thus might provide a better agreement between theory and 
observations. Whether associated with the envelope or the hot cores, our data make larger demands on the shock heating model, since the gas 
temperatures must be considerably higher. An open question is whether larger temperatures alone could reconcile the difference between 
the model predictions and measurements for the water vapor abundances in the Sgr~B2 envelope.

\subsection{Orion~KL}

Wilson et al. (1993) identified two NH$_3$ components toward Orion~KL, one with $T_{\rm rot}$ $\sim$ 165\,K, the other with $T_{\rm rot}$
$\sim$400\,K. Our upper line intensity limit to the ($J,K$)=(18,18) transition  is consistent with  $T_{\rm rot}$$<$500\,K (Table~2). Thus our 
data agree with previous results. We conclude that, unlike Sgr~B2, Orion~KL does not host a detectable extremely hot NH$_3$ component.

\section{Outlook}

We have observed the ($J,K$) = (18,18) absorption line of ammonia (NH$_3$), arising from levels $\sim$3130\,K above the ground state, definitely 
toward the compact H{\sc ii} region Sgr~B2~(M) and very likely toward Sgr~B2~(N). There was no detection toward Orion~KL. For the hot
gas component observed in Sgr~B2 there are two equally plausible scenarios, namely: (1) an extended low density envelope surrounding Sgr~B2~(M) 
and (N) or (2) much more compact dense clouds associated with the hot cores of the star forming region. To discriminate between these possibilities, 
VLA observations at $\lambda$=7\,mm are desirable. 

From the point of view of excitation, the ($J,K$) = (18,18) line of ammonia is at the same level as the lines of the fundamental (4.7$\mu$m) 
vibration-rotation band of CO. Both appear to trace average gas temperatures $\geq$1000\,K (for CO, see e.g. Geballe \& Garden 1987, 1990). Such 
temperatures are likely to originate in shocked gas. However, the fundamental CO band is seen in Orion but not in Sgr~B2, which is opposite to what
we obtained with the (18,18) line. This may be the case since the (18,18) line is only visible in absorption against the radio continuum of Sgr~B2, 
while the geometry of the Orion molecular cloud with its foreground radio continuum source is less favorable. Also, the hot gas in Sgr~B2 may be 
much more compact than that in the nearby Orion cloud, thus making the detection of the CO emission more difficult. Required shock velocities 
may also differ for NH$_3$ and CO (for CO, see Draine \& Roberge 1984). To further investigate a potential correlation between the two tracers 
of the hot molecular medium, a dedicated study of the $v$=1--0 transitions of CO in Sgr~B2 would be worthwhile.

The existence of an extended hot molecular envelope with more than several 100\,K around Sgr~B2 is unique among the star forming regions in 
the Galaxy. Is the presence of this envelope related to the exceptional conditions in the galactic center region? The molecular gas near the 
galactic center is characterized by comparatively warm temperatures and high pressures. Stellar densities tend to surpass those of the giant 
molecular clouds, and cloud-cloud collisions also having the potential to produce molecular debris may be more common than in other parts of the 
Galaxy. It is most likely that such gas is to be found in the centers of galaxies, but similar conditions in the disks of spiral galaxies cannot 
be excluded. Observing gravitational lenses amplifying a highly redshifted radio nucleus and jet may be the best method to find more such regions.

\end{document}